\documentclass[11pt]{article}
\usepackage{amsmath,amsthm,amssymb}
\usepackage{graphicx}
\usepackage{enumerate}
\theoremstyle{definition}
\newtheorem{rem*}{Remark}
\newtheorem*{rems*}{Remarks}
\newtheorem*{def*}{Definition}

\newcommand{\dd}{\mathrm{d}}
\newcommand{\ee}{\mathrm{e}}
\newcommand{\ii}{\mathrm{i}}
\renewcommand{\Im}{\operatorname{Im}}

\renewcommand{\Re}{\operatorname{Re}}

\DeclareMathOperator{\diag}{diag}

\begin{document}
\title{Initial boundary value problem 
for the focusing NLS equation with Robin boundary condition: half-line approach}
\author{Alexander Its$^1$ and Dmitry Shepelsky$^2$}
\date{}
\maketitle

\begin{abstract}
We consider the initial boundary value problem
for the focusing nonlinear Schr\"odinger equation in the
quarter plane $x>0,t>0$ in the case of decaying initial data (for $t=0$, as $x\to +\infty$)
and the Robin boundary condition at $x=0$. We revisit the approach based on the simultaneous
spectral analysis of the  Lax pair equations \cite{F02},  \cite{FIS05} and show that the method can be 
implemented without any a priori assumptions on the long-time behavior of the boundary values.
\end{abstract}

\section{Introduction}
\setcounter{equation}{0}

 The inverse scattering transform (IST) method
for studying initial value (IV) problems for certain nonlinear evolution
equations --- integrable nonlinear equations possessing a Lax pair representation ---
is known as a powerful tool for obtaining rigorous results concerning the most subtle
issues about the behavior of solutions of these problems, including detailed 
long-time behavior. The most efficient implementation of the IST method turns out to be based
on the Riemann--Hilbert (RH) problem method, which is essentially the reformulation of the 
scattering problem for one of the Lax pair equation ---
the $x$-equation --- in terms of an analytic (matrix) factorization problem
of the Riemann--Hilbert type.

The studies on the adaptation of the IST method to initial boundary value (IBV)
problems yield particular classes of boundary conditions, under which the 
IBV problem remains completely integrable, i.e. solving it reduces to solving a series of linear 
problems. At the beginning of 1990s it has been realized \cite{Sk, T88, F89, BT91, T91} that this is the 
case when the  boundary conditions allow an appropriate continuation,
based on the B\"acklund transformation,
 of the given initial data to the whole axis,
which  reduces the study of the IBV problem to the study of the associated IV problem.

In a recent paper \cite{DP10}, this  B\"acklund transformation approach is thoroughly revised
and adjusted to the modern, Riemann--Hilbert framework. This allowed the authors of \cite{DP10} 
to apply the nonlinear steepest descent method
\cite{DZ93}
 and analyze in great details several 
important asymptotic and stability questions  \cite{HZ09} in the theory of the focusing 
 nonlinear Schr\"odinger equation (NLS).

A general approach to IBV problems for integrable nonlinear
equations was originated in 1993 in  \cite{F93},   
and it has been actively developed since then \cite{FI94, FI96, BFS04, FIS05}.
In this approach, one treats the IBV problems with general
boundary conditions by using 
the spectral analysis of the both linear equations constituting the Lax pair.
The relevant Riemann-Hilbert problem is posed now on the union (cross) of  real and imaginary axis,
rather than on the real axis only as in the case of the IV problems. 
A key factor that affects the efficiency of the method is that the construction 
of the underlying Riemann--Hilbert problem
requires the knowledge of the spectral functions
that are determined, in general, by an excessive number of boundary values.
These boundary values cannot be
prescribed arbitrarily for a well-posed IBV problem and thus the problem of compatibility
of the boundary and initial values arises. It turns out that this compatibility
can be expressed, in a rather simple, explicit way, in terms of the associated spectral functions \cite{FIS05},
which in turn allows obtaining, particularly, the detailed asymptotic pictures.
But translating this description of compatibility into the physical space (of boundary and initial
values) requires solving, in general, nonlinear problems \cite{BFS03}, which makes the whole problem non-integrable.

However, for particular boundary conditions, an additional symmetry in the spectral space
allows bypassing the nonlinear step of resolving the compatibility issue and thus making 
the IBV problem as integrable as the associated IV problem \cite{F02} (see also \cite{BFS04, FIS05, FL10}).

In \cite{F02} (see also  \cite{FIS05}), the boundary conditions (called linearizable) leading to integrable
IBV for the  NLS equation have been specified by applying a symmetry 
analysis to the associated Lax pair equations. Particularly, the Dirichlet, Neumann, and Robin
boundary conditions were selected in this way, and the associated 
Riemann--Hilbert problem formalism  was presented,
under assumption that the boundary values being considered as a ``potential'' in the 
time-type equation from the Lax pair decay as $t\to \infty$ and, moreover, generate
the associated spectral functions with appropriate analytic properties. Furthermore,
in the paper \cite{FK04} it was shown that for the integrable boundary conditions  and, specifically,
for the Robin boundary condition, the original master RH problem posed
on the cross  can be unfolded to the
problem posed on the real line only. This, in principle, allows one to reproduce the
earlier results of  \cite{T91} within the general scheme of \cite{F93}.

The important issue which has not been fully addressed in \cite{F02},  \cite{FIS05}, and 
\cite{FK04} is the following. As it has been already mentioned, the original
master RH problem is formulated on the cross {\it under  a priory assumption}
that the boundary data rapidly decay as $t \to \infty$. On the other hand, using simple explicit soliton-like
solutions of the NLS equation, one can easily see that this assumption is not
necessarily true for solutions satisfying a linearizable boundary condition.  
Indeed, the two-parameter solutions 
$$
u(x,t)=\frac{\alpha}{\cosh(\alpha x+\phi_0)}\ee^{\ii \alpha^2 t}
$$
with $\alpha\in \mathbb R$ and $\phi_0\in \mathbb R$, which are obviously not decaying, 
for any $x$, as $t\to\infty$, satisfy the Robin condition
$u_x(0,t)+q\cdot u(0,t)=0$ with $q=\alpha\cdot\tanh{\phi_0}$.
Moreover, 
in the case $q>0$ the large-$t$ behavior of the solution is oscillatory for generic initial 
data (see the discussion in Section 4).
Hence the necessity
to produce an independent proof that the Riemann-Hilbert problem obtained
in \cite{FIS05} and \cite{FK04} for linearizable boundary conditions 
indeed yields  the solution of the IBV problem in question. In the present
paper, using some ideas which go back to the late 1980s works
on algebro-geometric  solutions of integrable PDEs, we show how this 
independent proof can be achieved and hence complete the program
originated in \cite{F02}.

We revisit the IBV for the  NLS equation with linearizable (Robin)
boundary condition
\begin{equation} \label{ibv}
\begin{aligned}
   & \ii u_t + u_{xx} + 2|u|^2 u =0,  & x>0, t>0, \\
& u(x,0)=u_0(x), & x\ge 0, \\
& u_x(0,t)+q u(0,t)=0, \quad & t\ge 0, 
 \end{aligned}
\end{equation}
where (a) $u_0(x)$ decays to $0$ as $x\to +\infty$ and (b) $q$ is a real constant.
Solving (\ref{ibv}) by the RH method consists of three steps: 
\begin{itemize}
\item[(i)]
provide a family of the Riemann--Hilbert 
problems parametrized by $x$ and $t$ such that the solution $u(x,t)$ of (\ref{ibv})
is expressed in terms of the solutions of these problems;
\item[(ii)] prove that  $u$ satisfies the initial condition $u(x,0)=u_0(x)$;
\item[(iii)] prove that $u$ satisfies the boundary condition $u_x(0,t)+q u(0,t)=0$.
\end{itemize}

Now let us comment on this procedure.
Concerning (i), the construction of the RH problem has to involve only the spectral functions 
associated with the initial condition $u_0(x)$. In the general framework of the simultaneous spectral 
analysis of the Lax pair equations, the RH problem is naturally formulated (see \cite{FIS05})
on the contour consisting of two lines: $k\in\mathbb{R}$ and  $k\in \ii\mathbb{R}$;
this reflects the fact that the spectrum of the $t$-equation from the Lax pair
with coefficients that are finitely supported or decaying at infinity consists of these lines.
But for particular boundary conditions, the contour can be deformed (two rays of the 
imaginary axis can be  fold down to the positive real axis) to the 
real axis only \cite{FK04}. Then it is the inherited symmetry property of the jump matrix 
for the deformed problem that allows verifying \emph{directly} that the boundary condition holds.
As for the verification of the initial condition, it is based on the fact that for $x=0$,
the original RH problem can be deformed in the opposite way (the rays of the 
imaginary axis are  fold down to the negative real axis) thus reducing the problem to that 
associated with
the $x$-equation of the Lax pair with the potential $u_0(x)$ \cite{FIS05}.

In this way we re-derive the results of \cite{T91}, \cite{DP10} and hence show that
the approach to 
 linearizable IBV problems stemming from 
the general methodology of simultaneous spectral analysis of the Lax pair equations \cite{F02},  \cite{FIS05}
can be implemented without any a priori assumptions about the long-time behavior
of the boundary values.

\section{The Riemann--Hilbert formalism for IBV problems}
\setcounter{equation}{0}

First, let us recall the Riemann--Hilbert formalism for 
IBV problems on the half-line $x\ge 0$ for the NLS equation \cite{FIS05}.

The focusing NLS equation 
\begin{equation}\label{nls}
\ii u_t + u_{xx} + 2|u|^2 u =0
\end{equation}
 is the compatibility condition
of two linear equations (Lax pair) \cite{ZS71}:
\begin{equation}\label{Lax1}
 \Psi_x+\ii k  \sigma_3 \Psi =U \Psi
\end{equation}
with 
\begin{equation}\label{U}
 U=\begin{pmatrix}
       0 & u \\ -\bar u & 0
      \end{pmatrix}
\end{equation}
and 
\begin{equation}\label{Lax2}
 \Psi_t+2\ii k^2  \sigma_3 \Psi =V \Psi
\end{equation}
with 
\begin{equation}\label{V}
 V=\begin{pmatrix}
       \ii |u|^2 & 2ku + \ii u_x \\ -2k\bar u+\ii \bar u_x & -\ii |u|^2
      \end{pmatrix}.
\end{equation}

Assuming that $u(x,t)$ satisfies (\ref{nls}) for $x>0$ and $0<t<T$ with some $T<\infty$,
define the solutions $\Psi_j(x,t,k)$, $j=1,2,3$ of (\ref{Lax1})--(\ref{V})
as follows: $\Psi_j(x,t,k)=\Phi_j(x,t,k)\ee^{(-\ii k x -2\ii k^2 t)\sigma_3}$,
where $\Phi_j$ solve the integral equations 
\begin{subequations}   \label{phi}
\begin{align}
\Phi_1(x,t,k)&=I -\ee^{-\ii kx\sigma_3}\int_t^T \ee^{-2\ii k^2 (t-\tau)\sigma_3}
    V(0,\tau,k) \Phi_1(0,\tau,k)\ee^{2\ii k^2 (t-\tau)\sigma_3}\ee^{\ii kx\sigma_3} \notag\\
& +\int_0^x \ee^{-\ii k(x-y)\sigma_3}
U(y,t)\Phi_1(y,t,k)\ee^{\ii k(x-y)\sigma_3}\dd y,\label{phi1}\\
\Phi_2(x,t,k)&= I +\ee^{-\ii kx\sigma_3}\int_0^t \ee^{-2\ii k^2 (t-\tau)\sigma_3}
    V(0,\tau,k) \Phi_2(0,\tau,k)\ee^{2\ii k^2 (t-\tau)\sigma_3}\ee^{\ii kx\sigma_3} \notag\\
& +\int_0^x \ee^{-\ii k(x-y)\sigma_3}
U(y,t)\Phi_2(y,t,k)\ee^{\ii k(x-y)\sigma_3}\dd y,\label{phi2}\\
\Phi_3(x,t,k)&=I-\int_x^{\infty}\ee^{-\ii k(x-y)\sigma_3}
U(y,t)\Phi_3(y,t,k)\ee^{\ii k(x-y)\sigma_3}\dd y.\label{phi3}
\end{align}
\end{subequations}

Define the scattering matrices $s(k)$ and $S(k)$, $k\in \mathbb{R}$,
 as the matrices relating the eigenfunctions
$\Phi_j(x,t,k)$ for all $x$ and $t$:
\begin{equation}\label{scat}
 \Psi_3(x,t,k)=\Psi_2(x,t,k)s(k),\qquad \Psi_1(x,t,k)=\Psi_2(x,t,k)S(k).
\end{equation}
The symmetry 
\begin{equation}\label{sym}
 \overline{\Psi_{11}(x,t,\bar k)}=\Psi_{22}(x,t,k),\qquad \overline{\Psi_{12}(x,t,\bar k)}=-\Psi_{21}(x,t,k)
\end{equation}
implies that 
\begin{equation}\label{s-S}
s(k)=\begin{pmatrix}
      \bar a(k) & b(k) \\ -\bar b(k) & a(k)
     \end{pmatrix},\qquad S(k)=\begin{pmatrix}
      \bar A(k) & B(k) \\ -\bar B(k) & A(k)
     \end{pmatrix}.
\end{equation}
From (\ref{phi}) and (\ref{scat}) it follows that 
the spectral functions $a(k)$ and $b(k)$ can be analytically continued into the upper
half-plane $k\in{\mathbb{C}}^+$ as bounded functions, with $a(k)\to 1 $ and $b(k)\to 0$
as $k\to\infty$.
Moreover, they are determined by $u(x,0)$, $x\ge 0$ only,
via 
$$
s(k)=(\Psi_2)^{-1}(x,0,k)\Psi_3(x,0,k).
$$

Similarly, the spectral functions $A=A(k;T)$ and $B=B(k;T)$ 
are entire functions bounded in the domains $I$ and $III$,
where $I=\{k: \Im k>0, \Re k >0\}$ and $III=\{k: \Im k<0, \Re k <0\}$,
 with $A(k;T)\to 1 $ and $B(k;T)\to 0$
as $k\to\infty$.
Moreover, they are determined by $u(0,t)$ and $u_x(0,t)$ for  $0\le t\le T$ only,
via 
$$
S(k;T)=(\Psi_2)^{-1}(0,t,k)\Psi_1(0,t,k).
$$

The compatibility of the set of functions $\{u(x,0),u(0,t),u_x(0,t)\}$ as traces
of a solution $u(x,t)$ of the NLS equation can be expressed in terms of the associated
spectral functions as follows:
\begin{equation}\label{gr}
 A(k;T)b(k)-a(k)B(k;t) = c(k;T)\ee^{4\ii k^2 T}, \qquad \Im k\ge 0,
\end{equation}
with some analytic $c(k;T)=O(\frac{1}{k})$ as $k\to\infty$
(in the general scheme \cite{F02} of analysis of IBV problems, (\ref{gr}) is called the
\emph{global relation}).

Define 
\begin{equation}\label{d}
 d(k)=a(k)\overline{A(\bar k)}+b(k)\overline{B(\bar k)}, \qquad k\in II=\{k:\Im k>0, \Re k <0\}.
\end{equation}
Finally, assuming  that $d(k)$
has at most a finite number of simple zeros in II and 
$a(k)$ has at most a finite number of simple zeros in ${\mathbb{C}}^+$,
define a piecewise meromorphic function (the superscripts denote the  column of the respective matrix)
\begin{equation}\label{M}
M(x,t,k)=\begin{cases}
           \begin{pmatrix}
            \dfrac{\Phi_2^{(1)}}{a}\ \  \Phi_3^{(2)}\end{pmatrix}, & \Im k>0, \Im k^2>0\\
     \begin{pmatrix} \dfrac{\Phi_1^{(1)}}{d}\ \  \Phi_3^{(2)}\end{pmatrix}, &
            \Im k>0, \Im k^2<0\\
  \begin{pmatrix} \Phi_3^{(1)} \ \  \dfrac{\Phi_1^{(2)}}{\bar d}\end{pmatrix}, &
            \Im k<0, \Im k^2>0\\
 \begin{pmatrix} \Phi_3^{(1)} \ \  \dfrac{\Phi_2^{(2)}}{\bar a}\end{pmatrix}, &
            \Im k<0, \Im k^2<0
          \end{cases}.
\end{equation}
Then the scattering relations (\ref{scat}) imply that the limiting
values of $M$ on the cross $\Im k^2=0$ satisfy the jump relations
\begin{equation}\label{M-jump}
 M_+(x,t,k)=M_-(x,t,k)\ee^{-\ii(kx+2k^2 t)\sigma_3}J_0(k)\ee^{\ii(kx+2k^2 t)\sigma_3},
\end{equation}
where
\begin{equation}\label{J0}
 J_0(k)=\begin{cases}
           \begin{pmatrix}
1+|r(k)|^2 & \bar r(k) \\ r(k) & 1
\end{pmatrix}, & k>0, \\
     \begin{pmatrix} 1 & 0 \\ \Gamma(k) & 1
\end{pmatrix}, &  k\in \ii\mathbb{R}_+,\\
  \begin{pmatrix} 1 & \bar \Gamma(\bar k) \\ 0 & 1 \end{pmatrix}, &
            k\in \ii\mathbb{R}_-,\\
 \begin{pmatrix} 1+|r(k)+\Gamma(k)|^2  & \bar r(k) + \bar \Gamma(k) \\
  r(k) + \Gamma(k) & 1
 \end{pmatrix}, & k<0,
          \end{cases}
\end{equation}
where $r(k)=\bar b(k)/a(k)$,
\begin{equation}\label{ga}
 \Gamma(k) = -\dfrac{\bar B(\bar k)}{a(k)d(k)}. 
\end{equation}
The orientation of the contour is chosen as from $-\infty$ to $+\infty$
along $\mathbb{R}$ and away from $0$ along $\ii\mathbb{R}$.

Complemented with the normalization condition $M=I+O(1/k)$ as $k\to\infty$
and the respective residue conditions at the zeros of $a(k)$ and $d(k)$
(see \cite{FIS05} for details),
the jump relation (\ref{M-jump}) can be viewed as the Riemann--Hilbert problem:
given $\{a(k),b(k),A(k),B(k)\}$, find $M(x,t,k)$ for all $x\ge 0$ and $t\ge 0$.
Then the solution of the NLS equation, $u(x,t)$, is given in terms of $M(x,t,k)$
by 
\begin{equation}\label{u-RHP}
 u(x,t)=2\ii \lim_{k\to\infty}k M_{12}(x,t,k).
\end{equation}
Moreover, $u(x,0)$ generates $\{a(k),b(k)\}$  and 
$\{u(0,t),u_x(0,t)\}$ generates $\{A(k),B(k)\}$ as the corresponding 
spectral functions provided the latter verify the global relation (\ref{gr}).
Therefore, the Riemann--Hilbert problem approach gives the solution
of the overdetermined IBV problem
\begin{equation} \label{ibv-g}
\begin{aligned}
  & \ii u_t + u_{xx} + 2|u|^2 u =0, \qquad & x>0, t>0,\\
& u(x,0)=u_0(x),& x\ge 0, \\
& u(0,t)=v_0(t), & 0\le t\le T,  \\
& u_x(0,t)=v_1(t), &0\le t\le T
\end{aligned}
\end{equation}
provided that the spectral functions $\{a(k),b(k),A(k),B(k)\}$
constructed from $\{u_0(x),v_0(t),v_1(t)\}$ satisfy the global relation (\ref{gr}).

\section{The Riemann--Hilbert formalism for Robin boundary condition}
\setcounter{equation}{0}

Even in  a conditional context presented in the previous Section, 
the RH method allows obtaining useful information
about the solution, e.g., that concerning the large-time behavior, see \cite{FIS05}.
However, there are cases when one can overcome the conditional nature of the solution
and solve a well-posed initial boundary value problem. The key factor \cite{F02}(see also  
\cite{FIS05})
for making this
possible is the existence of an additional symmetry in the spectral problem for the $t$-equation
of the Lax pair. 

This holds in the case of  Robin boundary condition.
Indeed, if $u+qu_x=0$ with some $q\in \mathbb{R}$, then the matrix 
$\tilde V:=V-2\ii k^2\sigma_3$ of the $t$-equation $\Psi_t=\tilde V \Psi$
satisfies the symmetry relation \cite{FIS05}
\begin{equation}\label{V-sym}
 \tilde V(x,t,-k)=N(k)\tilde V(x,t,k) N^{-1}(k),
\end{equation}
where $N(k)=\diag\{N_1(k), N_2(k)\}$ with $N_1(k)= 2k+\ii q$ and $N_2(k)= -2k+\ii q$.
In turn, (\ref{V-sym}) implies the symmetry for $S$: $S(-k;T)=N(k)S(k;t)N^{-1}(k)$,
which reads in terms of $A$ and $B$ as 
\begin{eqnarray}\label{AB-sym}
 A(-k;T)&=&A(k;T), \nonumber\\
B(-k;T)&=&-\frac{2k+\ii q}{2k-\ii q}B(k;T).
\end{eqnarray}

Now observe that the global relation (\ref{gr}) combined with the symmetry 
relation (\ref{AB-sym}) allows rewriting the RH problem (\ref{M-jump})
in the form that uses only the spectral functions $a(k)$ and $b(k)$
associated with the initial values $u(x,0)$. Indeed, since the exponential in the r.h.s.
of (\ref{gr}) is rapidly decaying for  $k\in I$,
the global relation (\ref{gr}) suggests to replace 
$\frac{B}{A}(k;T)$ by $\frac{b}{a}(k)$ for $k\in I$. Then, the symmetry 
relation (\ref{AB-sym}) suggests to replace $\frac{B}{A}(k;T)$ by 
$-\frac{2k-\ii q}{2k+\ii q}\frac{b}{a}(-k)$ for $k\in III$ and consequently
to replace $\frac{\bar B}{\bar A}(\bar k;T)$ by 
$-\frac{2k+\ii q}{2k-\ii q}\frac{\bar b}{\bar a}(-\bar k)$ for $k\in II$,
including the boundaries of respectively III and II. 
The resulting jump conditions have the same form as in (\ref{M-jump}),
\begin{equation}\label{M-t-jump}
 \tilde M_+(x,t,k)=\tilde M_-(x,t,k)\ee^{-\ii(kx+2k^2 t)\sigma_3}J_0(k)\ee^{\ii(kx+2k^2 t)\sigma_3},
\end{equation}
but with $\Gamma(k)$ replaced (cf. \cite{FIS05}) by
\begin{equation}\label{ga-t}
 \tilde\Gamma(k) = \frac{\overline{b(-\bar k)}}{a(k)}
\frac{2k+\ii q}{(2k-\ii q)a(k)\overline{a(-\bar k)}-(2k+\ii q)b(k)\overline{b(-\bar k)}}.
\end{equation}

Now we note that although $\Gamma(k)$ is defined, for general boundary values, 
only for $k\in II$, the function $\tilde \Gamma(k)$, in the generic case of absence of the
zeros of the denominator in (\ref{ga-t}), is analytic (and bounded) for all $k\in{\mathbb{C}}^+$
(and continuous up to the boundary). On the other hand, the exponentials 
in $\begin{pmatrix}
     1 & 0 \\ \tilde \Gamma(k) \ee^{2\ii k x + 4\ii k^2 t} & 1
    \end{pmatrix}
$
and 
$\begin{pmatrix}
     1 & \overline{\tilde \Gamma(\bar k)} \ee^{-2\ii k x - 4\ii k^2 t} \\ 0 & 1
    \end{pmatrix}
$
are bounded in respectively I and IV. Thus we can deform the RH problem with jump 
(\ref{M-t-jump}) on the cross to that on the real axis by defining
\begin{equation}\label{M-h}
 \hat M(x,t,k)=\begin{cases}
               \tilde  M(x,t,k), & k\in II\cup III, \\
\tilde M(x,t,k)\begin{pmatrix}
     1 & 0 \\ \tilde \Gamma(k) \ee^{2\ii k x + 4\ii k^2 t} & 1
    \end{pmatrix}, & k\in I, \\
\tilde M(x,t,k)\begin{pmatrix}
     1 & -\overline{\tilde \Gamma(\bar k)} \ee^{-2\ii k x - 4\ii k^2 t} & 1 \\ 0 & 1
    \end{pmatrix}, & k\in IV.
               \end{cases}
\end{equation}
The resulting jump conditions take the form
\begin{equation}\label{M-h-jump}
 \hat M_+(x,t,k)=\hat M_-(x,t,k)\ee^{-\ii(kx+2k^2 t)\sigma_3}\hat J_0(k)\ee^{\ii(kx+2k^2 t)\sigma_3},
\quad k\in \mathbb{R},
\end{equation}
where 
\begin{equation}\label{j-0-h}
 \hat J_0(k) = \begin{pmatrix}
                1+|r_e(k)|^2 & \bar r_e(k) \\ r_e(k) & 1
               \end{pmatrix}
\end{equation}
with 
\begin{equation}\label{r-e}
 r_e(k)=r(k)+\tilde \Gamma(k) = \frac{(2k-\ii q )\overline{b(k)}\overline{a(-k)}+
(2k+\ii q) \overline{b(-k)}\overline{a(k)}}{(2k-\ii q) a(k)\overline{a(-k)}-
(2k+\ii q) \overline{b(-k)}b(k)}.
\end{equation}
If the denominator in (\ref{r-e}) has zeros in  ${\mathbb{C}}^+$, then 
the formulation of the Riemann--Hilbert problem,
normalized by $\hat M\to I $ as $k\to\infty$,
is to be complemented by the 
residue conditions at these points (in this case, we make the genericity assumption that 
these zeros are simple and finite in number).

\begin{rem*} As it has already been mentioned in the introduction, 
the important observation that under the symmetry relations (\ref{AB-sym})
the RH problem on the cross can be deformed to the RH problem on the real axis
was first made in \cite{FK04}. 
\end{rem*}

\begin{rem*} We emphasize that the analytical continuation $\tilde\Gamma(k)$ of
the function $\Gamma(k)$ is not obliged to satisfy equation (\ref{ga}) on the
positive real axis  where, in view of the global relation,  it would have led
to the erroneous conclusion that  $r_{e}(k)$ must vanish for all positive  $k$.
\end{rem*}

The RH problem (\ref{M-h-jump}) coincides with the RH problems 
obtained in \cite{T91} and \cite{DP10}  via the B\"acklund technique
mentioned in the introduction. Our derivation, as being based
on the general IBV methodology,  is different. It also should be
noted that we have made certain a priory assumptions. Indeed,
to be able to push $T \to \infty$ in (\ref{gr}) and hence replace 
$\frac{B}{A}(k;T)$ by $\frac{b}{a}(k)$ for $k\in I$ we need to assume
a fast $t$-decay of the boundary data. The latter is not necessarily
true for the Robin boundary conditions (indeed, it is generally {\it not}
true). In what follows we show that, independently of the
previous considerations, the RH problem (\ref{M-h-jump})
yields indeed the solution of the NLS equation on the half line
with the Robin boundary condition. To this end, as it has already
been indicated in the Introduction, we shall show that the function
$u(x,t)$ generated by this RH problem satisfies (a) the NLS
equation, (b) the given initial conditions, and (c) (most challenging)
the Robin boundary condition.

The first part of the program is easy. 
The RH problems (\ref{M-t-jump}) and (\ref{M-h-jump}) both give the solution of the 
NLS equation in the domain $x>0$, $t>0$ via (\ref{u-RHP}); this is a standard fact
based on ideas of the dressing method, see, e.g., \cite{FT}. 

In order to verify the initial condition $u(x,0)=u_0(x)$, one observes that 
for $t=0$, the exponentials 
in $\begin{pmatrix}
     1 & 0 \\ \tilde \Gamma(k) \ee^{2\ii k x} & 1
    \end{pmatrix}
$
and 
$\begin{pmatrix}
     1 & \overline{\tilde \Gamma(\bar k)} \ee^{-2\ii k x} \\ 0 & 1
    \end{pmatrix}
$
are also bounded in respectively II and III. Thus we can deform the RH problem with jump 
(\ref{M-t-jump}) on the cross to that on the real axis by defining
\begin{equation}\label{M-c}
 \check M(x,t,k)=\begin{cases}
                \tilde M(x,t,k), & k\in I\cup IV, \\
\tilde M(x,t,k)\begin{pmatrix}
     1 & 0 \\ -\tilde \Gamma \ee^{2\ii k x } & 1
    \end{pmatrix}, & k\in II, \\
\tilde M(x,t,k)\begin{pmatrix}
     1 & \overline{\tilde \Gamma} \ee^{-2\ii k x } & 1 \\ 0 & 1
    \end{pmatrix}, & k\in III,
               \end{cases}
\end{equation}
which results in the  jump condition
\begin{equation}\label{M-c-jump}
 \check M_+(x,0,k)=\check M_-(x,0,k)\ee^{-\ii kx\sigma_3}\check J_0(k)
\ee^{\ii(kx\sigma_3},
\quad k\in \mathbb{R},
\end{equation}
where 
\begin{equation}\label{j-0-c}
 \check J_0(k) = \begin{pmatrix}
                1+|r(k)|^2 & \bar r(k) \\ r(k) & 1
               \end{pmatrix}.
\end{equation}
But the resulting RH problem (with residue conditions modified appropriately, see \cite{FIS05})
coincides with that for the spectral mapping $\{u_0(x)\} \mapsto \{a(k),b(k)\}$,
which yields $u(x,0)=u_0(x)$ due to the uniqueness of the solution of the RH problem.

The usefulness of the RH problem in the form (\ref{M-h-jump}) is that it allows
to verify that
$u(x,t)$ satisfies the Robin boundary condition by 
 using a symmetry of $r_e(k)$ followed from its construction (\ref{r-e}). This symmetry is as follows:
\begin{equation}\label{r-e-sym}
 r_e(-k) = r_e(k)\frac{\alpha(k)}{\overline{\alpha(k)}},
\end{equation}
where 
\begin{equation}\label{al}
 \alpha(k) = (2k-\ii q)a(k)\overline{a(-\bar k)} - (2k+\ii q)b(k)\overline{b(-\bar k)}.
\end{equation}
It is convenient to normalize $\alpha(k)$, which is 
analytic in ${\mathbb{C}}^+$, in such a way that it approaches $1$ as $k\to\infty$
and that it has neither a zero no a pole at $k=\frac{\ii|q|}{2}$.
Depending on the sign of $q$ and the behavior of $a(k)$ and $b(k)$ at $k=\frac{\ii|q|}{2}$,
different normalizing factors are needed. Indeed, if one introduces $a_e(k)$ and $\beta$ by
\begin{equation}\label{a-e}
 a_e(k)=\begin{cases}
         \frac{\alpha(k)}{2k-\ii q} = 
a(k)\overline{a(-\bar k)} - \frac{2k+\ii q}{2k-\ii q}b(k)\overline{b(-\bar k)}, &
	\text{if }\  q<0, a(-\frac{\ii q}{2})\ne 0 \ \text{or }\ q>0, b(\frac{\ii q}{2})=0 \\
\frac{\alpha(k)}{2k+\ii q} = \frac{2k-\ii q}{2k+\ii q}a(k)\overline{a(-\bar k)} 
	- b(k)\overline{b(-\bar k)}, &
	\text{if }\  q>0, b(\frac{\ii q}{2})\ne 0 \ \text{or }\ q<0, a(-\frac{\ii q}{2})=0
        \end{cases}
\end{equation}
and respectively 
\begin{equation}\label{be}
 \beta=\begin{cases}
         \frac{q}{2}, &
	\text{if }\  q<0, a(-\frac{\ii q}{2})\ne 0 \ \text{or }\ q>0, b(\frac{\ii q}{2})=0 \\
-\frac{q}{2}, &
	\text{if }\  q>0, b(\frac{\ii q}{2})\ne 0 \ \text{or }\ q<0, a(-\frac{\ii q}{2})=0
        \end{cases}
\end{equation}
 then
the requirements above are satisfied for $a_e(k)$ while the symmetry condition takes the form
\begin{equation}\label{r-e-sym1}
 r_e(-k) = r_e(k)\frac{a_e(k)}{\overline{a_e(k)}}\frac{k-\ii\beta}{k+\ii\beta}.
\end{equation}

The symmetry (\ref{r-e-sym1}) yields a certain $k \to -\bar{k}$
symmetry of the solution $\hat{M}(x,t,k)$ of the RH problem (\ref{M-h-jump}).
The relevant  symmetry relation has been established in \cite{DP10} in the
case $t = 0$ and $x\in \mathbb{R}$ . We shall need a version of that relation for the ``complimentary''
case, i.e. $x =0$, and $t>0$. We shall perform the derivation in this case 
following practically the same arguments as in \cite{DP10} (cf. the proof of Proposition 4.28
of  \cite{DP10}).

Denote  $\hat J(x,t,k)$ the jump matrix of problem (\ref{M-h-jump}), i.e.,
$$
\hat J(x,t,k)=\ee^{-\ii(kx+2k^2 t)\sigma_3}\hat J_0(k)\ee^{\ii(kx+2k^2 t)\sigma_3}.
$$
From (\ref{r-e-sym1}) it follows that
\begin{equation}\label{J-sym}
 \overline{\hat J(0,t,- k)} = \begin{pmatrix}
     \frac{1}{|a_e(k)|^2} & r_e(k)\frac{a_e(k)}{\overline{a_e(k)}}\frac{k-\ii\beta}{k+\ii\beta}
\ee^{4\ii k^2 t} \\
\overline{r_e(k)}\frac{\overline{a_e(k)}}{a_e(k)}\frac{k+\ii\beta}{k-\ii\beta}\ee^{-4\ii k^2 t} & 1
                              \end{pmatrix}
\equiv C_-(k) \hat J(0,t,k) C_+^{-1}(k),
\end{equation}
where 
\begin{equation}\label{C}
C(k)=\begin{cases}
      \begin{pmatrix}
       a_e(k) & 0 \\ 0 & \frac{1}{a_e(k)}
      \end{pmatrix}
\begin{pmatrix}
 k-\ii\beta & 0 \\ 0 & k+\ii\beta
\end{pmatrix}
\sigma_1, & \Im k >0,\\
\begin{pmatrix}
       \frac{1}{\overline{a_e(\bar k)}} & 0 \\ 0 & \overline{a_e(\bar k)}
      \end{pmatrix}
\begin{pmatrix}
 k-\ii\beta & 0 \\ 0 & k+\ii\beta
\end{pmatrix}
\sigma_1, & \Im k < 0,
\end{cases}
\end{equation}
with $\sigma_1 = \begin{pmatrix}
                   0 & 1 \\ 1 & 0
                  \end{pmatrix}
$.
This implies that the function 
\begin{equation}\label{M-b}
\breve M (t,k):={\cal E}(t,k)\overline{\hat M(0,t,-\bar k)}C(k)
\end{equation}
satisfies the same jump condition as $\hat M(0,t,k)$ does:
$$
\breve M_+(t,k)= \breve M_-(t,k)\hat J(0,t,k), \qquad k\in\mathbb{R}.
$$

The factor ${\cal E}(t,k)$ can be chosen so that $\breve M (t,k)\to I$
as $k\to\infty$ and that $\breve M (t,k)$ has no singularities at $k=\pm\ii\beta$.
Indeed,
introducing 
\begin{equation}\label{E}
{\cal E}(t,k)=\sigma_1 P(t)\begin{pmatrix}
                              \frac{1}{k-\ii\beta} & 0 \\ 0 &  \frac{1}{k+\ii\beta}
                             \end{pmatrix}
P^{-1}(t)
\end{equation}
with $P(t)$ to be defined, one has 
$$
\breve M (t,k) = \sigma_1 P \begin{pmatrix}
  \left(P^{-1}\overline{\hat M(0,t,-\bar k)}\right)_{11} & \left(P^{-1}\overline{\hat M(0,t,-\bar k)}\right)_{12}
\dfrac{k+\ii\beta}{k-\ii\beta} \\
\left(P^{-1}\overline{\hat M(0,t,-\bar k)}\right)_{21}
\dfrac{k-\ii\beta}{k+\ii\beta} & \left(P^{-1}\overline{\hat M(0,t,-\bar k)}\right)_{22}
                            \end{pmatrix}
D(k)\sigma_1,
$$
where 
\begin{equation}\label{D}
D(k)\equiv\diag\{d_1(k), d_2(k)\}=\begin{cases}
      \begin{pmatrix}
       a_e(k) & 0 \\ 0 & \frac{1}{a_e(k)}
      \end{pmatrix}
 & \Im k >0,\\
\begin{pmatrix}
       \frac{1}{\overline{a_e(\bar k)}} & 0 \\ 0 & \overline{a_e(\bar k)}
      \end{pmatrix}, & \Im k < 0.
\end{cases}
\end{equation}
Therefore, 
 there are no  singularities at $k=\pm\ii\beta$ provided
$$
\left(P^{-1}\overline{\hat M(0,t,\ii\beta)}\right)_{12}=0\ \  \text{and}\ \  
\left(P^{-1}\overline{\hat M(0,t,-\ii\beta)}\right)_{21}=0.
$$
This suggests to  determine $\bar P^{-1}(t)$ as follows:
$$
\bar P^{-1}(t) = \begin{pmatrix}
 \hat M_{22}(0,t,\ii\beta) & -\hat M_{12}(0,t,\ii\beta) \\
-\hat M_{21}(0,t,-\ii\beta) & \hat M_{11}(0,t,-\ii\beta)
                 \end{pmatrix},
$$
which gives
\begin{equation}\label{P}
 \bar P(t)=\frac{1}{\Delta(t)} \begin{pmatrix}
 \hat M_{11}(0,t,-\ii\beta) & \hat M_{12}(0,t,\ii\beta) \\
\hat M_{21}(0,t,-\ii\beta) & \hat M_{22}(0,t,\ii\beta)
                 \end{pmatrix}
\end{equation}
with 
\begin{eqnarray*}
\Delta(t) &=& \hat M_{11}(0,t,-\ii\beta)\hat M_{22}(0,t,\ii\beta)-\hat M_{12}(0,t,\ii\beta)\hat M_{21}(0,t,-\ii\beta )\\
&=& |\hat M_{11}(0,t,-\ii\beta)|^2 +|\hat M_{21}(0,t,-\ii\beta )|^2 >0.
\end{eqnarray*}

If the setting of the RH problem includes the poles, then, similar to the  case considered
in  \cite{DP10}, one can verify that $\breve M(t,k)$ defined by (\ref{M-b})--(\ref{P})
satisfies the same residue conditions as $\hat M(0,t,k)$
and hence the uniqueness of the solution of the RH problem gives
$$
\breve M(t,k)=\hat M(0,t,k),
$$
which reads in terms of $\hat M(0,t,k)$ only as 
\begin{equation}\label{M-sym}
 \overline{\hat M(0,t,-\bar k)} = \sigma_1 \bar P(t) \begin{pmatrix}
                              \frac{1}{k-\ii\beta} & 0 \\ 0 &  \frac{1}{k+\ii\beta}
                             \end{pmatrix}
\bar P^{-1}(t)\hat M(0,t,k)\begin{pmatrix}
 k-\ii\beta & 0 \\ 0 & k+\ii\beta
\end{pmatrix} D(k) \sigma_1.
\end{equation}
\begin{rem*} The above arguments allow actually to prove the general symmetry formula
which is valid for all $x$ and $t$,
$$
 \overline{\hat M(-x,t,-\bar k)} = \sigma_1 \bar P(t) \begin{pmatrix}
                              \frac{1}{k-\ii\beta} & 0 \\ 0 &  \frac{1}{k+\ii\beta}
                             \end{pmatrix}
\bar P^{-1}(t)\hat M(x,t,k)\begin{pmatrix}
 k-\ii\beta & 0 \\ 0 & k+\ii\beta
\end{pmatrix} D(k) \sigma_1
$$
In the case $t = 0$ and $x\in \mathbb{R}$, this is (up to the notations) the formula proven in   \cite{DP10}.
\end{rem*}

We shall now show  how the symmetry relation (\ref{M-sym}) alone  can be used to establish the Robin
boundary condition for $u(x,t)$. To this end we first evaluate, using   
(\ref{M-sym}), the entries $\hat M_{11}(0,t,-\ii\beta)$ and 
$\hat M_{21}(0,t,-\ii\beta)$. We have:
\begin{equation}\label{11}
\begin{aligned} 
\overline{\hat M_{11}(0,t,-\ii\beta)} & = \lim_{k\to -\ii\beta}\left(
\bar P(t) \begin{pmatrix}
                              \frac{1}{k-\ii\beta} & 0 \\ 0 &  \frac{1}{k+\ii\beta}
                             \end{pmatrix}
\bar P^{-1}(t)\hat M(0,t,k)\begin{pmatrix}
 k-\ii\beta & 0 \\ 0 & k+\ii\beta
\end{pmatrix} D(k)\right)_{22} \\
& = \bar P_{22}(t)\left(\bar P^{-1}(t)\hat M(0,t,-\ii\beta)\right)_{22} d_2(-\ii\beta) = 
\frac{1}{\Delta(t)}\hat M_{22}(0,t,\ii\beta)d_2(-\ii\beta) \\
& = \overline{\hat M_{11}(0,t,-\ii\beta)}\frac{d_2(-\ii\beta)}{\Delta(t)},
\end{aligned}
\end{equation}
where we have used the basic symmetry (\ref{sym}).
Similarly, we have 
\begin{equation}\label{21}
\begin{aligned} 
\overline{\hat M_{21}(0,t,-\ii\beta)} & = \lim_{k\to -\ii\beta}\left(
\bar P(t) \begin{pmatrix}
                              \frac{1}{k-\ii\beta} & 0 \\ 0 &  \frac{1}{k+\ii\beta}
                             \end{pmatrix}
\bar P^{-1}(t)\hat M(0,t,k)\begin{pmatrix}
 k-\ii\beta & 0 \\ 0 & k+\ii\beta
\end{pmatrix} D(k)\right)_{12} \\
& = \bar P_{12}(t)\left(\bar P^{-1}(t)\hat M(0,t,-\ii\beta)\right)_{22} d_2(-\ii\beta) = 
\frac{1}{\Delta(t)}\hat M_{12}(0,t,\ii\beta)d_2(-\ii\beta) \\
& = -\overline{\hat M_{21}(0,t,-\ii\beta)}\frac{d_2(-\ii\beta)}{\Delta(t)}.
\end{aligned}
\end{equation}

Comparing (\ref{11}) and (\ref{21}) gives
\begin{equation}\label{m-0}
\overline{\hat M_{11}(0,t,-\ii\beta)}(1-\theta)=0 \ \ \text{and}\ \ 
\overline{\hat M_{21}(0,t,-\ii\beta)}(1+\theta)=0,
\end{equation}
where
\begin{equation}\label{theta}
\theta = \frac{d_2(-\ii\beta)}{\Delta},
\end{equation}
which implies that either $\hat M_{11}(0,t,-\ii\beta)=0$ or 
$\hat M_{21}(0,t,-\ii\beta)=0$ for all $t\ge 0$. 

The next (and the last) step is to explore an idea that was first suggested 
by A. Bobenko in the late 1980s, and which was used then in several works
dealing with the algebro-geometric solutions of
integrable equations (see e.g. \cite{BI89}). 

Recall that $\Psi(t,k):=(\hat M_{11}(0,t, k)\ee^{-2\ii k^2 t}, 
\hat M_{21}(0,t, k)\ee^{-2\ii k^2 t})^T$
satisfies the differential equation (\ref{Lax2}) with $u=u(0,t)$ and $u_x=u_x(0,t)$,
i.e.
\begin{equation}\label{dif}
\begin{aligned} 
\frac{d\Psi_1}{dt} + 2\ii k^2 \Psi_1 & = \ii|u(0,t)|^2\Psi_1+ (2ku(0,t)+\ii u_x(0,t))\Psi_2, \\
\frac{d\Psi_2}{dt} - 2\ii k^2 \Psi_2 & = -\ii|u(0,t)|^2\Psi_2+ (-2k\bar u(0,t)+\ii \bar u_x(0,t))\Psi_1.
\end{aligned}
\end{equation}
From (\ref{dif}) it follows that 
 if $\Psi_1(t,-\ii\beta)=0$ for all $t\ge 0$ then $-2\ii\beta u(0,t)+\ii u_x(0,t)\equiv 0$,
or $u_x(0,t)-2\beta u(0,t)\equiv 0$;
 if $\Psi_2(t,-\ii\beta)=0$ then $2\ii\beta \bar u(0,t)+\ii \bar u_x(0,t)\equiv 0$,
or $u_x(0,t)+2\beta u(0,t)\equiv 0$. Observe that, according to (\ref{be}), $\beta$ 
can be either $q/2$ or $-q/2$.
But since the initial data satisfy the boundary condition $u_x(0,0)+qu(0,0)=0$,
by continuity it follows that this condition holds for all $t$.

A closer look at (\ref{m-0}) and (\ref{theta}) reveals that one can specify precisely whether
(a) $\hat M_{11}(0,t,-\ii\beta)=0$ or 
(b) $\hat M_{21}(0,t,-\ii\beta)=0$ occurs, depending on the sign of $q$ and the properties 
of $a(\ii|q|/2)$ and $b(\ii|q|/2)$. Indeed, since 
$\Delta=|\hat M_{11}(-\ii\beta)|^2+|\hat M_{21}(-\ii\beta)|^2>0$,
the choice between (a) and (b) is determined by the sign of $d_2(-\ii\beta)$.
According to (\ref{a-e}) and (\ref{be}), one can distinguish four cases.

(i) If $q>0$ and $b(\ii q/2)=0$, then $\beta=\frac{q}{2}>0$ and thus (see (\ref{D}))
$d_2(-\ii\beta)=\overline{a_e(\frac{\ii q}{2})}$. In turn, from (\ref{a-e}) it follows that 
in this case, $\overline{a_e(\frac{\ii q}{2})}=\left|a(\frac{\ii q}{2})\right|^2>0$ and thus
$1+\frac{d_2(-\ii\beta)}{\Delta}>0$, which implies (see(\ref{m-0})) that $\hat M_{21}(0,t,-\ii\beta)=0$.

(ii) If $q<0$ and $a(-\ii q/2)\ne 0$, then $\beta=\frac{q}{2}<0$ and thus 
$d_2(-\ii\beta)=\left(a_e(-\frac{\ii q}{2})\right)^{-1} = \left|a(-\frac{\ii q}{2})\right|^{-2}>0$.
Hence, in this case one also has $1+\frac{d_2(-\ii\beta)}{\Delta}>0$
and thus $\hat M_{21}(0,t,-\ii\beta)=0$.

(iii) If $q<0$ and $a(-\ii q/2)=0$, then $\beta=-\frac{q}{2}>0$ and thus
$d_2(-\ii\beta)= \overline{a_e(-\frac{\ii q}{2})} =-\left|b(-\frac{\ii q}{2})\right|^{2}<0 $.
Hence, in this case $1-\frac{d_2(-\ii\beta)}{\Delta}>0$, which implies that $\hat M_{11}(0,t,-\ii\beta)=0$.

(iv) If $q>0$ and $b(\ii q/2)\ne 0$, then $\beta=-\frac{q}{2}<0$ and thus
$d_2(-\ii\beta)=\left(a_e(\frac{\ii q}{2})\right)^{-1} = -\left|b(\frac{\ii q}{2})\right|^{-2}<0$.
Hence, in this case one also has $1-\frac{d_2(-\ii\beta)}{\Delta}>0$, which 
implies that $\hat M_{11}(0,t,-\ii\beta)=0$.

Summarizing, we see that $\hat M_{21}(0,t,-\ii\beta)=0$ corresponds to $\beta=\frac{q}{2}$
while $\hat M_{11}(0,t,-\ii\beta)=0$ corresponds to $\beta=-\frac{q}{2}$, which is indeed consistent 
with the fact that (\ref{dif}) implies $u_x(0,t)+qu(0,t)=0$ for all $t$.

\section{Concluding remarks}
\setcounter{equation}{0}

1. In the general Riemann--Hilbert approach to initial boundary value problems
for integrable nonlinear equations \cite{FI94, FI96, BFS04, FIS05},
an important step is the verification that the solution of the underlying nonlinear equation
obtained from the solution  of the associated RH problem indeed satisfies the prescribed 
boundary conditions. In the general case, this can be done by mapping the master RH problem,
in which the space parameter is taken to correspond to the boundary, to the RH problem 
associated with the $t$-equation of the Lax pair with a ``potential'' constructed from the prescribed
boundary values,
and showing that they are equivalent, in the sense that they produce
the same ``potentials''. But this means that 
the latter RH problem must be well-defined, which in particular requires, in the case of semi-infinite
time interval $0<t<\infty$, a precise description of the large-time behavior of the boundary
values. On the other hand, such description is, generally, not available in full
 from  the boundary conditions
of a well-posed initial boundary value problem, which forces one to make certain a priori assumptions about this behavior.

On the other hand, for linearizable boundary conditions, as we have shown on the example of the 
Robin boundary condition for the NLS equation, it is possible to verify \emph{directly}
that these boundary conditions hold, by using  additional symmetry properties of  an 
\emph{appropriately deformed} master RH problem thus avoiding restricting a priori assumptions.
The importance of this fact can be illustrated by the following simple observation:
in the case $q>0$, if  $b(\ii q/2)\ne 0$ (generic case!), then, as it follows from  (\ref{a-e}),
 $a_e(\ii q/2) = -|b(\ii q/2)|^2 <0$. On the other hand, $a_e(\ii\xi)\to 1$ as $\xi\to +\infty$.
Noticing that $a_e(\ii\xi)\in {\mathbb R}$ for all $\xi>0$, we conclude that 
 $a_e(k)$ must have at least one  zero for $k\in \ii{\mathbb{R}}_+$. Consequently,  
as it follows from the asymptotic analysis similar to the one 
done in \cite{DP10}, stationary solitons generated by these zeros 
dominate the large-$t$ asymptotics in the direction 
 along the $t$-axis, which prevents from  assuming the decaying
behavior of $u(0,t)$ and $u_x(0,t)$ needed for defining the associated spectral functions
$A(k)$ and $B(k)$.

2.
For the initial data that do not produce zeros of $a_e(\ii\xi)$ and thus stationary solitons,   
the boundary value of the solution of the Robin problem decays as $t^{-1/2}$,
as it again follows form the asymptotic analysis similar to \cite{DP10}. 
This is formally not enough in order to proceed 
with the approach of \cite{F02},  \cite{FIS05}.
However, one can guarantee the needed 
rate of decay of the boundary value assuming
that certain combination of the spectral functions $a(k)$ and $b(k)$, 
corresponding to the initial data $u_0(x)$ in the IST formalism, vanishes at $k =0$ to a high order.

3. In our analysis of the IBV problem formulated for the domain $x\ge 0$, $t\ge 0$,
we do not go ``beyond the domain'', working with $x$ and $t$, as parameters in the RH problem,
that stay in the  domain prescribed by the problem. 
This is in contrast with the approach based on an appropriate
continuation to $x<0$ with the help of  the B\"acklund transformation   allowing to control the necessary conditions at $x=0$ for all $t$  \cite{BT91, T88, T91, DP10}.
Actually, folding the contour of the RH problem down to the real axis establishes
the relationship between these approaches. Indeed, the functions 
$r_e(k)$ and $a_e(k)$ are exactly the reflection coefficient and the inverse transmission
coefficient for the RHP problem on the whole axis associated with
the initial value  problem on the whole line with the  B\"acklund-continued initial data,
see \cite{DP10}.

4. An alternative way to see that $u(x,t)$ constructed from the solution of the RH problem
satisfies the boundary condition is based on using the further terms in the expansion 
of $\hat M(x,t,k)$ as $k\to\infty$ giving not only $u(x,t)$ (as in (\ref{u-RHP})) but also
$u_x(x,t)$ \cite{FIS05}:
\begin{equation}\label{u-x-RHP}
u_x(x,t)=\lim_{k\to\infty}\left[4(k^2\hat M)_{12}(x,t,k) + 2\ii u(x,t) (k\hat M)_{22}(x,t,k)\right]
\end{equation}
Indeed, let us substitute the expansion (for $x=0$) 
$$
\hat M(0,t,k) = I + \frac{m^1(t)}{k} + \frac{m^2(t)}{k^2} + \dots, \qquad k\to\infty
$$
into the symmetry relation (\ref{M-sym}) taking into account that, in view of (\ref{a-e}) and (\ref{be}),
$$
a_e(k)=\begin{cases} 
        1+O(\frac{1}{k^2}), & \text{if}\  \beta=\frac{q}{2}, \\
	1-\frac{\ii q}{k} +O(\frac{1}{k^2}), & \text{if}\   \beta=-\frac{q}{2}
       \end{cases}
$$
Equating the terms of order $k^{-1}$ one gets
\begin{equation}\label{m-1-1}
 \begin{pmatrix}
  -2\overline{m^1_{11}} & 0 \\ 0 & -2\overline{m^1_{22}} 
\end{pmatrix}= \ii \beta \sigma_1
\bar P \sigma_3 \bar P^{-1}+\frac{\ii q}{2}\sigma_3.
 \end{equation}
On the other hand, from (\ref{P}) we have
$$
\bar P \sigma_3 \bar P^{-1} =\frac{1}{\Delta}\begin{pmatrix}
   \hat M_{11}(-\ii\beta) & \hat M_{12}(\ii\beta)  \\
 \hat M_{21}(-\ii\beta) & \hat M_{22}(\ii\beta) 
                                             \end{pmatrix}
\sigma_3
\begin{pmatrix}
   \hat M_{22}(\ii\beta) & -\hat M_{12}(\ii\beta)  \\
 -\hat M_{21}(-\ii\beta) & \hat M_{11}(-\ii\beta) 
                                             \end{pmatrix}
=\frac{1}{\Delta}\begin{pmatrix}
   \delta & 0 \\ 0 & -\delta
                 \end{pmatrix},
$$
where 
$ \delta = \hat M_{11}(-\ii\beta)\hat M_{22}(\ii\beta)+\hat M_{21}(-\ii\beta)\hat M_{12}(\ii\beta)$;
here the fact that either $\hat M_{11}(-\ii\beta)=0$ or  $\hat M_{21}(-\ii\beta)=0$ implies the diagonal
structure of the resulting matrix.

Further, in the case $\hat M_{11}(-\ii\beta)=0$ (recall that in this case $\beta=-\frac{q}{2} $) one has 
$$
\delta = \hat M_{21}(-\ii\beta)\hat M_{12}(\ii\beta) = -\left|\hat M_{21}(-\ii\beta)\right|^2,\quad
\Delta = \left|\hat M_{21}(-\ii\beta)\right|^2
$$
and thus $\bar P \sigma_3 \bar P^{-1}=-\sigma_3$.
In the case $\hat M_{21}(-\ii\beta)=0$ ($\beta=\frac{q}{2} $) one has 
$$
\delta = \hat M_{11}(-\ii\beta)\hat M_{22}(\ii\beta) = \left|\hat M_{11}(-\ii\beta)\right|^2,\quad
\Delta = \left|\hat M_{11}(-\ii\beta)\right|^2
$$
and thus $\bar P \sigma_3 \bar P^{-1}=\sigma_3$.
In both cases, from (\ref{m-1-1}) we have
\begin{equation}\label{m-1}
 m^1_{11}(t)=m^1_{22}(t) = 0.
 \end{equation}

Now, equating the terms of order $k^{-2}$, we have for the off-diagonal part:
$$
\begin{pmatrix}
 0 & -2 m^2_{21} \\ -2 m^2_{12} & 0 \\
\end{pmatrix}= \begin{pmatrix}
0 & m^1_{21}(-2m^1_{22}-\ii q) \\
m^1_{12}(-2m^1_{11}+\ii q) & 0
\end{pmatrix},
$$
which, in view of (\ref{m-1}), reads
\begin{equation}\label{m-12}
 m^2_{12}(t) = m^1_{12}(t)\left(-\frac{\ii q}{2}\right).
\end{equation}
Finally, (\ref{u-x-RHP}) and (\ref{u-RHP}), in view of (\ref{m-1})
and (\ref{m-12}), yield
$$
u_x(0,t) = 4m^2_{12}(t)=-2\ii q m^1_{12}(t) = -q u(0,t).
$$

\section*{Acknowledgments}

The work of A.I. was
supported in part by NSF Grant
DMS-1001777.
D.Sh. gratefully acknowledges the 
hospitality of the Courant Institute of Mathematical Sciences and the
Indiana University-Purdue University Indianapolis,
where part
of this research was done.

$^1$Indiana University-Purdue University Indianapolis\\
402 N.~Blackford St., Indianapolis, IN 46202-3267, USA\\
itsa@math.iupui.edu

\medskip

$^2$B.~Verkin Institute for Low Temperature Physics and Engineering\\
47 Lenin Avenue, 61103 Kharkiv, Ukraine\\
shepelsky@yahoo.com

\end{document}